\documentclass[twocolumn,superscriptaddress,prb]{revtex4}
\usepackage{graphicx}
\usepackage{epsfig}
\usepackage{amsmath, amssymb}
\usepackage{verbatim}
\usepackage{bm}

\newcommand{\gammabar}{\ensuremath\gamma\kern-0.53em-}

\begin{document}

\title{Continuous transitions between composite Fermi liquid and Landau Fermi
  liquid: \\ a route to fractionalized Mott insulators}
\author{Maissam Barkeshli}
\affiliation{Department of Physics, Stanford University, Stanford, CA 94305 }
\author{John McGreevy}
\affiliation{Department of Physics, Massachusetts Institute of Technology,
Cambridge, MA 02139 }

\begin{abstract}
One of the most successful theories of a non-Fermi liquid metallic state is the composite Fermi liquid (CFL) theory
of the half-filled Landau level. In this paper, we study continuous quantum phase transitions out of the CFL
state and into a Landau Fermi liquid, in the limit of no disorder and fixed particle number. This 
transition can be induced by tuning the bandwidth of the Landau level
relative to the interaction energy, for instance through an externally applied periodic potential.
We find a transition to the Landau Fermi liquid through a gapless Mott insulator with a Fermi 
surface of neutral fermionic excitations. In the presence of spatial symmetries, we also 
find a direct continuous transition between the CFL and the Landau Fermi liquid. 
The transitions have a number of characteristic observable signatures,
including the presence of two crossover temperature scales, resistivity jumps, and vanishing compressibility. 
When the composite fermions are paired instead, our results imply quantum critical points
between various non-Abelian topological states, including the $\nu = 1/2$ Moore-Read Pfaffian (Ising $\times$ U(1) topological order), 
a version of the Kitaev B phase (Ising topological order), and paired electronic superconductors. 
To study such transitions, we use a projective construction of the CFL, which goes beyond the conventional framework of flux 
attachment to include a broader set of quantum fluctuations.
These considerations suggest a possible route to fractionalized Mott insulators by starting with FQH states and 
tuning the Landau level bandwidth.
\end{abstract}

\maketitle

\section{Introduction}

Despite decades of work, the breakdown of Landau Fermi liquid theory in metallic states still poses some of the most challenging,
unsolved problems in condensed matter physics. This breakdown often occurs in the vicinity of quantum phase transitions
in Fermi liquid metals, although there are some situations, such as in the half-filled Landau level in
two-dimensional electron gases (2DEGs), where entire non-Fermi liquid metallic phases have been found to exist. Perhaps the 
most experimentally successful theory of any non-Fermi liquid metal in more than one dimension is the composite
Fermi liquid (CFL) theory of the half-filled Landau level, which has had a number of striking theoretical predictions 
that have been experimentally verified in GaAs 2DEGs.\cite{CompositeFermBook,compositeFermJain,HL9312,KZ9289,R9459}  
The CFL also provides structural insight: as the magnetic field is tuned,
the conventional series of fractional quantum Hall (FQH) plateaus in the lowest Landau level in GaAs can be
understood as integer quantum Hall (IQH) states of the composite fermions \cite{J8999,CompositeFermBook,compositeFermJain,HL9312,KZ9289,R9459}. 

The crucial reason that the half-filled Landau level gives rise to this distinct non-Fermi liquid state of electrons is that the 
Landau level has essentially zero bandwidth, thus quenching the kinetic energy and leaving
the interaction energy to dominate. As the bandwidth is increased to be on the order of the 
interaction strength, the system should pass through a quantum phase transition. In the limit that the 
bandwidth is large compared to the interaction strength, the resulting state will be well-described by
Landau Fermi liquid theory. This raises the question of whether bandwidth-tuned transitions out of quantum Hall states
and into more conventional states can be continuous, and if so, what the possible critical theories are. 
For example, can there be a continuous quantum phase transition out of the CFL state and into a Landau Fermi liquid in a clean system?
Such a continuous phase transition between a non-Fermi liquid metal and a Fermi liquid metal would be quite exotic; 
starting from the Landau Fermi liquid side, it would describe the continuous destruction of the 
electron Fermi surface and the emergence of a Fermi surface of composite fermions with
singular gauge interactions. Previous work on the fate of {\it incompressible} (Abelian) FQH states 
includes \Ref{KL9223,CF9349, WW9301, YS9809, S9857, W0050, BW1004, BM1293}.
Our analysis below will allow us to consider also the fate of incompressible non-Abelian FQH states, such as the 
Moore-Read Pfaffian state\cite{MR9162}, as the bandwidth is increased. 

One possible way to experimentally induce such bandwidth-tuned transitions is by imposing a periodic potential on a 2DEG subjected
to an external magnetic field. When there is $2\pi p/q$ flux per unit cell of a weak periodic potential, each Landau level 
splits into $p$ subbands, where the bandwidths and gaps between the subbands are on the order of the strength of the periodic potential. For
$2\pi$ flux per plaquette, the periodic potential does not split the bands, but only gives a bandwidth to the Landau levels and may be used to reach
the regime in which the bandwidth is comparable to, or much larger than, the interaction strength. There are potentially
other physical realizations as well. Recent attention has focussed on
bands with non-zero Chern number in lattice systems without an external magnetic field \cite
{NS11106, TW11106, SG1189, RB11, V1106, WG11, XLQ11, MS1101, MS1169, LR1234}. Partially filling such 
Chern bands can lead to various fractional quantum Hall (FQH) states. The existence of incompressible 
FQH states has already been numerically established, and it is natural to expect that the CFL state can also be realized in such situations. 
The bandwidth of the Chern band can be increased by applying pressure, eventually resulting in a Landau Fermi liquid and providing
another possible physical realization of the transitions discussed in this paper. In the context of the partially filled Chern bands,
there is no background magnetic field, so the conventional framework of flux attachment and flux-smearing mean-field theory
is inapplicable, raising a further fundamental conceptual question of how to understand the effective field theory of the CFL in such a situation.\cite{MS1101} 
\begin{figure}[t]
\centerline{
\includegraphics[width=3.3in]{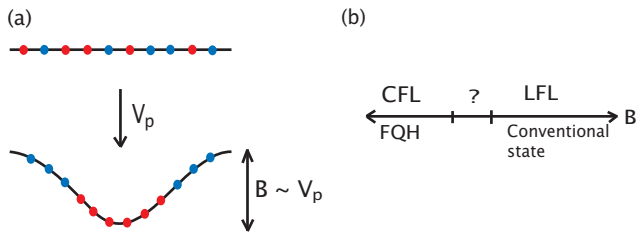}
}
\caption{(a) Schematic illustration of the effect of a periodic potential $V_p$ with $2\pi$ flux per plaquette on a half-filled Landau level.
The Landau level acquires a bandwidth (B) on the order of $V_p$; in the limit that $V_p$ is much larger than the interaction energy, the system
should be well-described by Landau Fermi liquid theory. (b) As the bandwidth (B) of the flat band is tuned, the system will transition 
out of the CFL and ultimately into the Landau Fermi liquid. More generally, the system will transition out of a fractionalized
QH state and ultimately into a more conventional state. 
\label{llBandwidth}
}
\end{figure}

In this paper, we address the questions discussed above. We begin by studying a lesser known, alternative formulation of the theory of the
CFL state using the projective construction.\footnote{See e.g. \Ref{AM0903} for a recent appearance in the literature.} 
This construction has several advantages over the conventional flux attachment approach.
Most importantly it allows us to include a broader range of quantum fluctuations,
which provides access to nearby states, including ones in which the gauge fluctuations are destroyed to 
yield more conventional electronic states. We use this construction to develop the CFL theory for partially filled Chern bands 
without an external magnetic field, where the conventional notion of flux attachment may not be applicable in general. 
Subsequently, we study several possible states that can be described within the same low energy effective theory. 
For gapless states, these include the composite Fermi liquid, a gapless Mott insulator (GMI) with a Fermi surface of emergent neutral
fermions, and the Landau Fermi liquid. When the fermions are paired instead, the CFL and GMI descend into 
either paired composite fermion states, such as the $\nu = 1/2$ Moore-Read (MR) Pfaffian, or 
gapped topological states with an emergent $Z_2$ gauge field coupled to the fermions. The Landau Fermi liquid is then replaced
by a paired electronic superconductor. These considerations then allow us to study zero temperature phase transitions 
out of the quantum Hall states and into other exotic fractionalized states, such as the GMI or its paired descendants, 
and ultimately into conventional electronic states, such as Landau Fermi liquids and superconductors (Fig. \ref{llBandwidth}). 

We find a transition between the CFL and the Landau FL through an intervening gapless Mott insulator state with emergent
Fermi surface. In the presence of some spatial symmetry, such as inversion, we find a quantum critical point directly separating
the CFL and the Landau FL. When we consider $p_x + ip_y$ pairing of the composite fermions, these considerations lead
to a transition between the $\nu = 1/2$ MR Pfaffian \cite{MR9162} and a $p_x +i p_y$ paired superconductor of electrons
through an intervening non-Abelian topological state that is topologically equivalent to the non-Abelian B phase of Kitaev's honeycomb model\cite{K0602}. 
In the presence of some spatial symmetry, our results then imply a quantum critical point directly separating the
$\nu=1/2$ MR Pfaffian and the conventional $p_x + ip_y$ electronic superconductor; in the presence of a strong magnetic field this 
may be induced by tuning a periodic potential with $2\pi$ flux per plaquette. 

Recently, building on a slave-rotor mean-field theory,\cite{FG0414} T. Senthil has developed a theory of a continuous Mott 
transition between a $U(1)$ spin liquid Mott insulator with a spinon Fermi surface and a Landau FL by including the effects of the $U(1)$ gauge 
fluctuations \cite{S0809}. That theory describes how the Fermi surface is continuously destroyed
as the transition is approached, and it contains a number of striking, experimentally testable predictions. These include
the existence of two crossover temperature scales and therefore two distinct quantum critical regimes, a universal resistivity 
jump at the transition, and diverging quasiparticle effective masses and Landau parameters as the transition is approached 
from the Fermi liquid side.  It has been conjectured\cite{M0505,LL0503} that such a gapless spin liquid state may be realized in a number
of different compounds,\cite{SM0301,HM0704,ON0707} which would then provide the possibility of observing such a bandwidth-tuned continuous Mott
transition.

The transitions between the CFL, GMI, and Landau Fermi liquid display phenomenology similar to that found in the $U(1)$ spin liquid Mott transition of Ref.~\onlinecite{S0809}.
We find that they can be continuous, and there are two crossover temperature scales and resistivity jumps at the transitions. 
Remarkably, we find that although the CFL and Landau FL are both compressible states, the quantum
critical point between them is incompressible at zero temperature. On the composite Fermi liquid side of the transitions, the second crossover temperature scale does not exist in the presence of long-range Coulomb interactions, but can be made
to appear by adding a metallic gate to screen the long-range interactions. 

While the compounds studied in Refs.~\onlinecite{SM0301,HM0704,ON0707} provide experimentally promising venues for the observation of such a continuous Mott transition,
our work suggests another experimentally promising venue for similar physics. If the CFL theory is taken seriously as describing
the half-filled Landau level, then our results predict that there is a GMI state nearby with a Fermi surface of neutral fermions and it
may be realized by tuning appropriate periodic potentials with a wavelength on the order of the interparticle spacing; 
this state is a non-trivial spinless analog of the $U(1)$ spin liquid Mott insulator with spinon Fermi surface. 
We may refer to this as a $U(1)$ orbital liquid Mott insulator. 
The proximity to a Landau Fermi liquid by tuning the bandwidth would then also allow for another experimentally 
promising venue to study the continuous destruction of the Fermi liquid and the appearance
of a fractionalized state with a Fermi surface coupled to an emergent $U(1)$ gauge field. 

This paper is organized as follows. In Section \ref{hlrRev}, we begin with a brief review of the construction of the 
CFL theory of the half-filled Landau level, and we review some of the results of the Halperin-Lee-Read theory
of the gauge fluctuations of this state. In Section \ref{PartonHLR}, we develop a projective/parton construction for the 
CFL state, and we show how it can be applied to situations without an external magnetic field, and can be generalized to any
filling fraction, including odd-denominator fillings. In Section \ref{neighborSec}, we discuss states proximate to the
CFL at half-filling, including the GMI with emergent Fermi surface, and the Landau Fermi liquid. In Section \ref{contTranSec},
we study the continuous transitions separating these states. In Section \ref{pairedSec} we discuss the consequences 
of this theory for various paired non-Abelian states.

\section{Brief review of the CFL theory}
\label{hlrRev}

The CFL theory of the compressible FQH state at $\nu = 1/2m$ begins by performing an exact transformation by which $2m$ units of
flux quanta are attached to each electron. Similar flux transmutations have been used to derive effective field theories 
for a variety of QH states.\cite{ZH8982,Z9225,CompositeFermBook,compositeFermJain} In the Lagrangian formulation, 
in the imaginary time formalism, we have
\begin{align}
Z = \int \mathcal{D} f^\dagger \mathcal{D} f \mathcal{D} a e^{-\int_0^\beta d\tau d^2 r \mathcal{L}},
\end{align}
with $\beta  = 1/T$, $T$ is the temperature, and the Lagrangian density is
\begin{align}
\label{cflLag}
\mathcal{L} = &f^\dagger (\partial_\tau - ia_0 - \mu) f -\frac{1}{2m_b} f^\dagger (\partial - i A - ia)^2 f 
\nonumber \\
&+ \frac{1}{2} \int d^2 r' V(r - r') f^\dagger(r) f^\dagger(r') f(r') f(r)
\nonumber \\
&+ \frac{i}{4\pi (2m)} \epsilon^{\mu \nu \lambda} a_\mu \partial_\nu a_\lambda .
\end{align}
$f$ is the composite fermion field and $A$ is the background electromagnetic field. The interaction may be chosen to be of the form
$V(r) \sim 1/r^\eta$, where $\eta = 1$ corresponds to the case of Coulomb interactions. If the fluctuations of 
$a$ are treated exactly, this is an exact transformation of the original theory. The value of this rewriting of the original theory
is that it allows for a mean-field approximation that was not available before. Since the filling fraction is $\nu = \frac{N_e}{N_\Phi} = \frac{1}{2m}$, 
there are $2m$ flux quanta for each electron. Since we have also added $2m$ units of $a$ flux to each electron, we can consider a mean-field
state where $\langle a \rangle = - A$. With this mean-field approximation, the composite fermions $f$ on average do not feel 
any magnetic field.

It is found that the theory above describes a compressible,\footnote{By finite compressibility, it is meant that 
the zero frequency density-density correlation function $\chi_{\rho \rho}(q)$ is finite as the wave vector 
$q \rightarrow 0$ for short-range interactions. For long-range interactions $V(q)$,
finite compressibility means that $\chi_{\rho \rho}(q) \sim 1/V(q)$.} metallic state, which describes well the phenomenology that is 
experimentally observed in the half-filled Landau level.\cite{CompositeFermBook}

The single-particle properties of $f$ exhibit various infrared singularities.\cite{HL9312} For example, the self-energy $\Sigma_f(\omega)$ of $f$ has a leading
singularity that, for Coulomb interactions ($\eta = 1$), behaves as
\begin{align}
\Sigma_f(\omega) \sim \omega \ln (i\omega)~.
\end{align}
For shorter-range interactions, where $1 < \eta \leq 2$, 
the self-energy gives rise to stronger singularities:
\begin{align}
\Sigma_f(\omega) \sim (i\omega)^{\frac{2}{\eta + 1}}
\end{align}
However, since $f$ is not gauge-invariant, its single-particle correlation functions cannot be directly measured. Instead, the physical
observables are the gauge-invariant response functions, which are found to be non-singular and Fermi liquid-like \cite{KF9417}. 

The electron operator, $c(r)$, in this theory is described as
\begin{align}
c(r) = \hat{M}^{2m}(r) f(r),
\end{align}
where $\hat{M}(r)$ is an instanton operator for the gauge field $a$ that annihilates $2\pi$ units of flux. The electron
is simply the composite fermion $f$, together with $2m$ units of flux of the $a$ gauge field. Using this electron operator, the 
equal-space electron Green's function was computed in \Ref{KW9478} using a semi-classical approximation for the
instanton action, with the result:
\begin{align}
G_+(\tau) \equiv \langle c(0, \tau) c^\dagger (0,0) \rangle \approx G_0(\tau) e^{-S_{M\bar{M}}(\tau) },
\end{align}
where $G_0(\tau)$ is an algebraically decaying function of $\tau$ and $S_{M\bar{M}} \propto \tau^s$,
where the exponent $s$ depends on the form of the interactions between the composite fermions. It was found that
the spectral function, $A_+(\omega)$, defined as the inverse Laplace transform of $G_+(\tau)$, behaves like
\begin{align}
\label{spectral}
A_+(\omega) \sim e^{- \alpha /\omega^\beta},
\end{align}
where $\alpha$ and $\beta$ again depend on the interactions between the composite fermions. This indicates a strong 
exponential suppresion of the tunneling density of states at low frequencies (see also \Ref{HP7793} for a different derivation with the
same conclusion). 

In the presence of long-range Coulomb interactions ($\eta  = 1$),  the specific heat was calculated to scale as\cite{HL9312}
\begin{align}
C_v \sim T \ln T.
\end{align}
With short-range interactions, the specific heat instead scales as
\begin{align}
C_v \sim T^{2/3}.
\end{align}
The behavior of the specific heat and the exponentially decaying spectral function are both signatures of strongly 
non-Fermi liquid behavior. 

A wave function for the CFL state that is expected to capture its long-wavelength properties is of the form \cite{RR9400}:
\begin{align}
\Psi(\{ r_i \}) = \mathcal{P}_{LLL} \bigl( (z_i - z_j)^{2m} \Det[e^{i k_i r_j} ] \bigr),
\end{align}
where $\mathcal{P}_{LLL}$ indicates the projection to the lowest Landau level. The factor $(z_i - z_j)^{2m}$ can be thought of
as attaching $2m$ units of flux quanta to the composite fermions, which are filling a Fermi sea. 

The above theory has been quite successful in explaining many 
long-wavelength phenomena observed experimentally at $\nu = 1/2$. Nevertheless, the above formulation has three 
shortcomings that we address in this paper\footnote
{An orthogonal set of shortcomings 
of a more microscopic nature is addressed in \eg~\Ref{R949, PH9871, MS9801, L9845, SH9912,S9627}.}. 
First, it yields a limited framework for understanding continuous phase transitions out of the
state. While this formulation is useful for studying transitions of the composite fermion Fermi 
surface, such as the pairing instability that leads to the Moore-Read Pfaffian state, more general continuous transitions, such as 
into a conventional Fermi liquid, cannot be understood through the above construction. 

The second shortcoming is that while this flux attachment procedure and the associated flux-smearing 
mean-field theory can be defined in models where the Chern bands are induced by an external magnetic field, 
it is in general unclear how to extend this to lattice models without an external magnetic field (see \Ref{MS1101} 
for a recent discussion). In fact, in many cases, the flux attachment procedure for a half-filled Chern band appears to 
fail entirely.

Thirdly, the above formulation of the composite particle theories on a compact space is problematic \cite{LF9923}:
because the Chern-Simons level is not quantized, the partition sum on a surface of genus $\geq 1$ is not 
invariant under large gauge transformations \cite{W89121, WN9077}.

In order to address these shortcomings, in the following section we develop a theory of the CFL through 
a projective construction.  

\section{Projective construction of the CFL state}
\label{PartonHLR}

Here we will study a derivation of the CFL through a totally different approach that does not begin from
notions of flux attachment and flux smearing. Instead, we use a projective/parton construction,
which provides several crucial advantages: most importantly,  it incorporates a broader set of
quantum fluctuations, which yields a path towards understanding transitions out of the CFL and 
into, for instance, Fermi liquids. Additionally, this formulation can be extended to lattice models without 
an external magnetic field, where the conventional flux attachment picture fails; it yields insight 
into the CFL state and its topological properties; and finally it suggests generalizations of the CFL state
to odd-denominator filling fractions or to states at the same filling fraction but which differ in their topological 
properties. Such projective constructions are more familiar in the study of quantum spin liquids\cite{wen04}, although they have
been developed both for non-Abelian FQH states\cite{W9102, W9927, BW1002,BW1121}, and more conventional 
Abelian states\cite{J8979,W9211,MS1169,LR1234,V1106}.

For concreteness, let us consider a system of spinless interacting fermions on a square lattice in the presence of a 
background external magnetic field:
\begin{align}
\label{eq:origH}
H = \sum_{ij} [t_{ij} c_i^\dagger c_j e^{i A_{ij}} - (\mu + A^0_i ) \delta_{ij} n_i + V_{ij} n_i n_j ],
\end{align}
with
\begin{align}
n_i = c_i^\dagger c_i.
\end{align}
Let $\phi = 2\pi p/q$ be the amount of flux of $A$ 
per plaquette. In the absence of interactions, the above model is then 
a tight-binding model with $q$ bands; in the limit $q \rightarrow \infty$, the lowest bands become flat and equally
spaced in energy, corresponding to Landau levels. We will suppose that the average density of electrons is such that the
lowest Landau level has a filling $\nu = 1/2m$. 

We begin by performing an exact rewriting of the above model by decomposing $c_i$ in terms of different bosonic and fermionic variables:
\begin{align}
\label{bf}
c_i = b_i f_i,
\end{align}
where $b_i$ annihilates a boson carrying the electric charge, and $f_i$ annihilates a neutral fermion. This introduces a $U(1)$ gauge symmetry
associated with the local transformation
\begin{align}
b_i \rightarrow e^{i\theta_i} b_i, \;\;\; f_i \rightarrow e^{-i \theta_i} f_i,
\end{align}
which keeps the electron operator invariant. We see:
\begin{align}
n_i = n_i^b n_i^f,
\end{align}
where $n_i^b = b_i^\dagger b_i$ and $n_i^f = f_i^\dagger f_i$. The states at each site are now labelled as
$|n^b_i, n^f_i \rangle$. These states form an expanded Hilbert space; the physical states in the expanded Hilbert space
are the gauge-invariant ones: $|0,0\rangle$ and $|1,1\rangle$, which correspond to the state with zero or one electron(s), respectively. 
We see that the physical states satisfy the constraint $n^b = n^f$, so 
\begin{align}
n_i = n^b_i= n^f_i.
\end{align}
Inserting this into the original Hamiltonian \eqref{eq:origH} gives:
\begin{align}
H = \sum_{ij} [t_{ij} b_i^\dagger b_j e^{i A_{ij}} f_i^\dagger f_j  - (\mu +A^0_i)\delta_{ij} n_i^b + V_{ij} n_i^b n_j^b].
\end{align}
Now we can decouple the quartic terms in $H$ using a self-consistent mean-field approximation:
\begin{align}
b_i^\dagger b_j e^{i A_{ij}} f_i^\dagger f_j \approx 
\frac{1}{2}[ \chi_{ij} b_i^\dagger b_j e^{i A_{ij}}  + \eta_{ij} f_i^\dagger f_j 
+ \chi_{ij} \eta_{ij}],
\end{align}
where
\begin{align}
\chi_{ij} &= \langle f_i^\dagger f_j \rangle, \;\;\;\;\;
\eta_{ij} = e^{i A_{ij}} \langle b_i^\dagger b_j \rangle. 
\end{align}
In the mean-field ansatz, we will assume that the ground state is such that $b$ forms a $\nu = 1/2m$ Laughlin
FQH state while $f$ forms a Fermi surface. This implies that $\eta_{ij}$ and $\chi_{ij}$ are both real-valued
and translationally invariant. It is simple to verify that such a mean-field state can be self-consistent, because 
if $\eta_{ij}$ is real-valued, then $f$ forms a Fermi surface with no average magnetic field, and 
its real-space averages will be real, which will imply that $\chi_{ij}$ is real. Similarly, if $\chi_{ij}$ is real, 
then $b$ feels a magnetic field around a plaquette set by $A_{ij}$, and therefore $\eta_{ij}$ can be real. 

The fluctuations about the mean-field state can be included in the manner 
familiar from parton constructions of other states: the amplitude fluctuations of $\chi$ and $\eta$ are
gapped, so we will only include the phase fluctuations. This leads to the following Hamiltonian:
\begin{align}
\label{mfHam}
H = \sum_{ij} [&t_{ij} (\chi_{ij} b_i^\dagger b_j e^{i A_{ij} + i a_{ij}} + \eta_{ij} e^{-i a_{ij}} f_i^\dagger f_j)
\nonumber \\
&- (\mu +A^0_i) n_i^b + a^0_i(n_i^b - n_i^f) + V_{ij} n_i^b n_j^b].
\end{align}
$a^0_i$ is introduced as a Lagrange multiplier to enforce the constraint that $n_i^b = n_i^f$.
While we derived this effective theory from a mean-field ansatz and included the allowed long-wavelength fluctuations, it can also be derived
somewhat differently, following the analysis of \Ref{LL0503} for the $U(1)$ spin liquid, as a saddle-point of 
the path integral by introducing additional fields to exactly decouple the quartic terms. 
Such mean-field approximations can yield stable deconfined fixed points of the resulting gauge theory.
Whether they correspond to a global minimum in the energy is a more detailed question 
of energetics that must be answered through numerical studies or by comparing the phenomenology of these 
theories with experimental observations. 

In a $\nu = 1/2m$ Laughlin state, the insertion of $2m$ units of flux quanta at sufficiently long wavelengths 
will create a single unit of charge. Therefore, at energies well below the gap of the bosonic Laughlin state, the boson can be represented by
the instanton operator:
\begin{align}
b = \hat{M}^{2m},
\end{align}
where $\hat{M}$ is the instanton operator which annihilates $2\pi$ flux of $a$. Therefore, at energies well below the gap of the bosonic state, 
the electron operator can be represented as
\begin{align}
\label{eq:electronop}
c = \hat{M}^{2m} f,
\end{align}
just as in the conventional CFL theory. In particular, this means that at low energies, the boson number density is
\begin{align}
\label{nb1}
n^b = \frac{1}{2\pi} \frac{1}{2m} \epsilon^{ij} \partial_i a_j,
\end{align}
which means that the boson interaction term in the Hamiltonian (\ref{mfHam}) can be rewritten in terms of $a$.

We can describe the $\nu = 1/2m$ bosonic Laughlin state by using a $U(1)_{2m}$ CS theory associated with a second 
emergent gauge field\cite{wen04}. Then, the effective theory becomes, to lowest order in a continuum approximation,
\begin{align}
\label{partonCFLLag}
\mathcal{L} &= \mathcal{L}_b + \mathcal{L}_f + \mathcal{L}_{int},
\nonumber \\
\mathcal{L}_b &= \frac{2m}{4\pi} \epsilon^{\mu \nu \lambda} \tilde{a}_\mu \partial_\nu \tilde{a}_\lambda +
\frac{1}{2\pi} \epsilon^{\mu \nu \lambda} (a_\mu + A_\mu) \partial_\nu \tilde{a}_\lambda,
\nonumber \\
\mathcal{L}_f &=  f^\dagger (i\partial_t + a_0) f + \frac{1}{2m^*} f^\dagger (\partial + i a)^2 f,
\nonumber \\
\mathcal{L}_{int} &= \int \frac{d^2 r'}{(4\pi m)^2} V(r-r') (\epsilon^{ij} \partial_i a_j(r)) (\epsilon^{ab} \partial_a a_b(r')).
\end{align}
The boson current $j_b^\mu$ is given in terms of $\tilde{a}$:
\begin{align}
j_b^\mu = \frac{1}{2\pi} \epsilon^{\mu \nu \lambda} \partial_\nu \tilde{a}_\lambda. 
\end{align}
Note this is consistent with (\ref{nb1}), because integrating out $\tilde{a}_0$ in (\ref{partonCFLLag}) in the absence
of $A$ yields the constraint $\epsilon_{ij} \partial_i a_j = 2m \epsilon^{ij} \partial_i \tilde{a}_j$.

Relabeling $a \rightarrow -(a + A)$ and subsequently integrating out $\tilde{a}$ yields the following Lagrangian 
density:\footnote{Note that integrating out $\tilde{a}$ this way is inconsistent on a genus $g \geq 1$ manifold and leads
to a gauge anomaly associated with large gauge transformations, as there are multiple topological sectors that must 
be properly considered. \cite{W89121, WN9077,LF9923}}
\begin{align}
\mathcal{L} = &f^\dagger (i\partial_t - a_0 - A_0) f + \frac{1}{2m} f^\dagger (\partial - i A -i a)^2 f 
\nonumber \\
&+ \frac{1}{2} \frac{1}{4\pi m} \epsilon^{\mu \nu \lambda} a_\mu \partial_\nu a_\lambda + \cdots
\nonumber \\
&+ \int \frac{d^2 r'}{(4\pi m)^2} V(r-r') (\epsilon^{ij} \partial_i a_j(r)) (\epsilon^{ab} \partial_a a_b(r')),
\end{align}
which is precisely the conventional CFL Lagrangian (\ref{cflLag}). It immediately follows that this construction yields a state
with all of the same thermodynamic and transport properties of the conventional CFL state. 
A point of subtlety is that this construction yields a compressible state, despite the fact that $b$ 
forms an incompressible Laughlin FQH state. Formally, this is possible because the polarizability of
the electrons, $\Pi_e$, is given by the Ioffe-Larkin sum rule \cite{IL8988}
\begin{align}
\label{ioffeLarkin}
\Pi_{e}^{-1} = \Pi_{b}^{-1} + \Pi_f^{-1},
\end{align}
where $\Pi_b$ and $\Pi_f$ are the polarizability tensors of the $b$ and $f$ particles, respectively. 
If the polarization tensors were diagonal, the inverse compressibilities and inverse conductivities
of the slave particles would add to give the inverse compressibility or inverse resistivity of the electron system. 
However, in this case, the polarization tensors have off-diagonal components, $\Pi_{0i} \neq 0$; thus, the compressibility of the electron system can be finite, despite the incompressibility of the boson sector. 
Less formally, the coupling to the gapless gauge flux of $a$ allows the Laughlin state
of the bosons to respond continuously to the introduction of extra particles.

The single-electron Green's function is
\begin{align}
G_e(r-r', t-t') &= \langle T c(r,t) c^\dagger(r', t') \rangle 
\nonumber \\
&= \langle T b(r,t) b^\dagger(r', t') f(r,t) f^\dagger(r', t') \rangle.
\end{align}
In the slave-particle mean-field approximation, where we ignore gauge fluctuations, 
\begin{align}
G^{mf}_e(r, t) = G_b(r,t) G_f(r,t)~.
\end{align}
Since $b$ forms a gapped FQH state, its bulk correlations will clearly decay exponentially in space and time, and
therefore so will $G^{mf}_e(r,t)$. 
The result of \Ref{KW9478} is recovered by considering gauge fluctuations. At low energies, we can 
represent $b$ by the instanton operator, so that the electron operator is given by (\ref{eq:electronop}). 
Now the calculation of the electron Green's function reduces to that of \Ref{KW9478}.

This mean-field theory of the CFL state can actually be derived at any filling fraction $\nu$ by allowing the bosons to 
form more generic incompressible Abelian or non-Abelian FQH states and is therefore not limited to even-denominator 
fillings. Even staying at $\nu=1/2$, it is possible to consider distinct incompressible states at $\nu = 1/2$ for the bosons instead of
the $1/2$ Laughlin state. A possible example is the $\nu =1/2$ bosonic orbifold state studied in \Ref{BW1121}. 
These states would have different topological orders in the boson sector, but would yield the same low energy dynamics described by
(\ref{cflLag}) if the bosons are integrated out. The different possible topological orders for the bosonic sector
suggest that the CFL should be viewed as a \it topological non-Fermi liquid\rm. An open conceptual problem is to develop
tools to characterize the topological order in such topological non-Fermi liquid metals.

\subsection{CFL in the absence of an external magnetic field: partially-filled Chern bands
and the `factorization' of bandstructure}

Recently, there has been wide interest in studying FQH states in lattice models without an external magnetic field,
but with partially filled flat bands with non-trivial topology. Since a flat band with Chern number 1 is topologically equivalent
to a Landau level, it is expected that when such a band is partially filled, it is possible to recover conventional
FQH states. So far, the focus has mostly been on the incompressible FQH states, however one may expect that
the CFL can also appear in such systems. This raises the question of how to construct the 
groundstate wavefunction and low-energy effective description 
for such a state.
It is simple to see that the conventional flux attachment and flux-smearing mean-field approximation fails for such
a situation.\footnote{We would like to thank Sri Raghu for useful conversations regarding this point} 
For example, consider a square lattice with a half-filled flat band with Chern number 1. Since the band is at half-filling,
this implies that there is one electron for every two sites. In the flux attachment picture, we must attach a multiple of $4 \pi$ flux
to obtain composite fermions. However, attaching $4\pi$ flux to each electron is equivalent to adding $2\pi$ flux per plaquette.
In the flux-smearing mean-field approximation, $2\pi$ flux per plaquette is equivalent to zero flux per plaquette, so the
statistical transmutation 
appears to be ineffective.

In contrast, the projective construction of the previous section can straightforwardly be generalized to the case of a partially
filled Chern band without an external magnetic field. 
In Refs.~\onlinecite{MS1169, LR1234}, a prescription was given for a 
mean field theory with the parton decomposition $ c = f_1 f_2...f_n $; 
the idea is that the hopping elements for the $f$s should be $n$th roots
of those of $c$.  
This was justified by a strong coupling expansion of the parton lattice gauge theory.
Here we wish to extend this to the parton decomposition $ c = fb$;
the hopping elements for $f$ and $b$ must then be {\it factors} of 
those of $c$, but need not be equal.

Consider a tight-binding model defined by the following hamiltonian
\begin{equation}
 H = - \sum_{rr' \in \mathcal{E}} t^c_{rr'} c^\dagger_{r} c_r' + {\text{h.c.}} + \text{interactions}.
 \end{equation}
Here $c^\dagger, c$ denote creation and annihilation operators
for electrons; 
$\mathcal{E}$ denotes links in the lattice.
Our mean field ansatz for the 
``factorization" of this lattice is
\begin{equation}
 H_{\mbox{mf}} = 
-  \sum_{rr'\in \mathcal{E}} \left(  t^f_{rr'} f^\dagger_{r} f_{r'} 
+ t^b_{rr'}b^\dagger_{r} b_{r'} + {\text{h.c.}}  \right)
 \end{equation}
where we demand that for each link
\begin{equation}
 t^c_{rr'} = t^f_{rr'} t^b_{rr'} h_{rr'}^{-1} ~.
\end{equation}
The scale $h_{rr'}$ is a parameter of the parton lattice gauge theory
representing the energy cost for allowing gauge flux along the link $rr'$;
the argument from strong coupling expansion proceeds as in \Ref{MS1169},
to which we refer the reader for the details.
The phases in $t^c$ which make the resulting bands topological
must be distributed between $t^f$ and $t^b$ -- this choice
is a lattice analog of the choice in the continuum parton construction of where to assign the electric charge 
amongst the partons.
The analog of letting the bosons carry the charge is 
to put all the phases in $t^b$.

\def\({\left(}
\def\){\right)}
\def\a{\alpha}
\def\b{\beta}

For definiteness, consider an electron model at quarter-filling on the checkerboard lattice.
Some relatively-flat Chern bands are realized by the following tight-binding model 
with next-nearest neighbor interactions \cite{SY0911}.
The Hamiltonian is $H = \sum_{k \in BZ}  \( h_{2N} + h_{3N}\)$ with
\begin{align}
h_{2N} &= - t e^{ i \varphi} \left[ \b^\dagger_k \a_k \( e^{ {i \over 2} \( k_x + k_y\) }
+  e^{ {i \over 2} \( - k_x + k_y\) } \)   \right. &\\
& \left.   \a^\dagger_k \b_k
\( e^{ {i \over 2} \( k_x - k_y\) } + e^{ {i \over 2} \( -k_x + k_y\) }
\)  \right] + {\text{h.c.}}   &
\nonumber
\\
\nonumber
h_{3N} &= - \a^\dagger_k \a_k \left(t_1'   e^{i k_x} + t_2' e^{i k_y} \right)
- \b^\dagger_k \b_k \left( t_2'  e^{i k_x} + t_1' e^{i k_y} \right)
\end{align}
where $\a$ and $\b$ are fourier modes of the electron creation operators on the two sublattices:
\begin{equation}
  \a_k = \frac{1}{\sqrt{N_a}} \sum_{x_a} e^{ik x_a} c_{x_a} , ~~~
  \b_k = \frac{1}{\sqrt{N_b}} \sum_{x_b} e^{ik x_b} c_{x_b} ~~,
\end{equation}
The bands may be further flattened by the addition of third-neighbor interactions \cite{SG11106}.
At quarter-filling, the electrons fill the lowest band halfway; this band has Chern number unity.

To construct a lattice analog of the HLR state, we take the ansatz:
\begin{equation}
t^b = t e^{i \phi}, ~ \(t_2'\)^b = - \(t_1'\)^b, ~~~
t^f = t, ~ \(t_2'\)^f = + \(t_1'\)^f~~.
\end{equation}
This puts the fermion in a state which half-fills a topologically-trivial band,
and puts the boson in a state which half-fills the lowest topologically non-trivial band with Chern number one.
Note that unlike the case of 
\Ref{MS1169}, no enlargement of the unit cell was required so far;
if we wanted to explicitly describe the lattice analog of the $\nu=1/2$ Laughlin
wavefunction of this boson, we would have to use a second layer of 
parton construction which in which the unit cell is doubled.

\section{States proximate to CFL: gapless Mott insulator and Landau Fermi liquid }
\label{neighborSec}

The main utility of the projective construction of the CFL state in the previous section is that now, within the same theory, we 
can find saddle points associated with other many-body states. 

For example, consider the $\nu = 1/2$ Landau level problem and suppose that we add a periodic potential, $V_p$, with a multiple of $2\pi$ flux per plaquette. 
$V_p$ couples to the boson density (see \eqref{mfHam}), and therefore will increase the bandwidth of the Landau level for $b$. When $V_p$ is on the
order of the boson gap, the bosons can undergo a transition into either a Mott insulating state for $b$ or a superfluid.\cite{BM1293}
Alternatively, in the case of the partially filled Chern band without an external magnetic field, we can consider increasing the pressure
to increase the hopping matrix elements. This can also cause transitions out of the bosonic Laughlin state and into the MI or the superfluid.

It is also possible to consider fractionalized Mott insulators for $b$, such as $Z_2$ topologically ordered states. 
In this paper, we will ignore these more complicated scenarios, as they are expected to be less likely. 
Instead we will focus on the simplest possible states for the bosons and the correspondingly simplest ways of 
transitioning out of the CFL and into the LFL. 


\subsection{Gapless Mott Insulator}

When the strength of the periodic potential is on the order of the interaction strength, we expect the bosons to 
transition out of the $\nu = 1/2$ Laughlin state. A simple, generic possibility is that the bosons continuously transition into
a trivial Mott insulator without any topological order. For this to occur, the bosons must be at integer filling per unit cell;
this can happen by either explicit or spontaneous translation symmetry breaking.

From the constraint $n_b = n_f$, integer filling of the bosons implies integer filling of the fermions. 
This means that the fermions can generally form either a band insulator or Fermi surfaces with equal
area of particles and holes. In the mean-field Hamiltonian, the fermion hopping is given by
\begin{align}
t_{ij}^f \propto \langle b_i^\dagger b_j \rangle.
\end{align}
Deep in the Mott insulating phase, it is possible that $\langle n^b_i \rangle \approx 1$ or $0$, depending on the lattice site.
Similarly, it is possible that $\langle b_i^\dagger b_j \rangle \approx 0$ for $i \neq j$. In this limit, the fermions cannot 
form a Fermi surface; they are localized to the sites where the bosons are located, and do not disperse. 

Closer to the $\nu = 1/2$ Laughlin and superfluid state, $\langle n^b_i \rangle$ is approximately uniform,
and the bosons have a larger correlation length, so $t_{ij}^f$ can be appreciable. In such a situation, the 
mean-field state of the fermions can be such that they form a Fermi surface. Since the fermions are at integer filling, the
area of hole and particle pockets must be equal. This state is closely related to the gapless $U(1)$ spin liquid 
Mott insulators with spinon Fermi surface and an emergent $U(1)$ gauge field.\cite{M0505,LL0503} The difference is that the gapless 
Mott insulator considered here consists of spinless electrons. While there is an integer number of electrons per 
unit cell, each unit cell consists of multiple sites; therefore, there is a pseudospin index that plays the role of spin 
and which can allow the system to have gapless degrees of freedom, despite the charge being localized and gapped. 

Starting from the CFL, where the fermions form a Fermi surface, the introduction of a periodic potential will fold
the Brillouin zone; if the Fermi surface has no nesting, as in the conventional scenarios, there 
will continue to be stable particle/hole pockets for the $f$ fermions. As the bosons undergo a continuous 
transition to the MI state, the $f$ sector changes gradually, as there is generically no change of spatial symmetry 
at the boson FQH -MI transition. Within the MI state, as the bosons are taken deeper into the MI and become fully 
localized, the Fermi surface can gradually shrink to zero and eventually the $f$ fermions will form a band insulator. 

Physically, we can understand the appearance of the GMI state by analogy with the CFL state. In the CFL state, the electron
fractionalizes into a boson $b$, which carries the electric charge, and the neutral $f$ fermion. The boson $b$ then forms
an incompressible Laughlin FQH state in order to minimize the strong interaction energy and the phase frustration
induced by the magnetic field, while $f$ is free to form a Fermi surface.
It is therefore reasonable to imagine that instead of $b$ forming a Laughlin FQH state, it will form a Mott insulator
in order to minimize the strong repulsive interaction energies. The many-body wave function that
describes such a state is the usual projected wave function \cite{wen04}:
\begin{align}
\label{gmiwfn}
\Psi_e(\{ r_i \}) &= \langle 0| \prod_i c(r_i) |\Phi_{mf} \rangle
\nonumber \\
&= \Psi_{b}(\{r_i\}) \Psi_{f}(\{r_i\}),
\end{align}
where $|\Phi_{mf} \rangle$ is the mean-field ground state of $b$ and $f$, where $f$ is forming a Fermi surface and
$b$ is forming a Mott insulator. This is equivalent to taking the boson MI wave function and fermion FS wave function
and projecting the bosons and fermions to the same location.

The effective theory for this state takes the form
\begin{align}
\mathcal{L} = \mathcal{L}_b + \mathcal{L}_f + \mathcal{L}_{a},
\end{align}
where $\mathcal{L}_b$ describes the gapped bosonic excitations, 
$\mathcal{L}_f$ describes the fermions in a Fermi sea and coupled to $a$, and 
$\mathcal{L}_a = \frac{1}{g^2} (\nabla \times a)^2 + \cdots$ is the
action for the $U(1)$ gauge field $a$. When the bosons are in the gapped, trivial MI state, 
they can be integrated out to give, to lowest order,
\begin{align}
\mathcal{L} = \mathcal{L}_f + \frac{1}{g^2}(\nabla \times a)^2.
\end{align}
The gapless Mott insulator is electrically insulating, thermally conducting, and incompressible. 
As a result, its existence cannot be established through DC electrical transport
measurements. Instead, it can be distinguished from the trivial Mott insulator by probing thermal
behavior, such as specific heat or thermal conductivity. The specific heat and thermal conductivity 
of such a state scale as $C_v \sim T^{2/3}$, and $K/T \sim T^{2/3}$,
respectively.\cite{LN9221, AI9448, L0902, MM1004,MS1007}
The possibility of thermal conduction in this spinless Mott insulator comes from the fact that
the thermal conductivity of the electrons is equal to the sum of the thermal conductivity of 
the $b$ and $f$ systems;\cite{LN9221} this is in contrast to the electrical transport, where typically it is
the resistivities that add.

The Fermi surface can be more directly probed through Friedel oscillations, as 
proposed recently in \Ref{MS1104} for the $U(1)$ spin liquid Mott insulator; in both cases, the density-density
correlation function has algebraic correlations displaying signatures of a Fermi surface. 

\subsection{Landau Fermi liquid}

When the interaction energy is small compared to the bandwidth, then it will be preferable for $b$ to 
condense into the bottom of its band and form a superfluid. 
In such a case, the emergent $U(1)$ gauge symmetry associated with 
$a$ is spontaneously broken, and the resulting state of the electrons is described by Landau Fermi liquid theory. 
In the Landau Fermi liquid, the quasiparticle residue $Z \sim |\langle b \rangle |^2$.

Since time-reversal symmetry is broken, with interactions the boson superfluid state will in general consist of
a normal component with non-vanishing orbital currents. The electron state will therefore be a Landau Fermi liquid with 
orbital loop currents and a non-zero Hall conductance, as time-reversal symmetry is explicitly broken. 

If the electrons are at integer filling per unit cell of the periodic potential, 
then the resulting state is either a band insulator or a Landau Fermi liquid with equal
area for electron and hole pockets. Within our construction, if we start from a CFL with non-nested Fermi surface
and add a periodic potential, the metallic case will be the generic situation. 

If the electrons are at fractional filling, then the resulting state is a Landau Fermi liquid
with no constraints on electron and hole pockets.

\section{Continuous Transitions}
\label{contTranSec}

\begin{figure}
\centerline{
\includegraphics[width=3.3in]{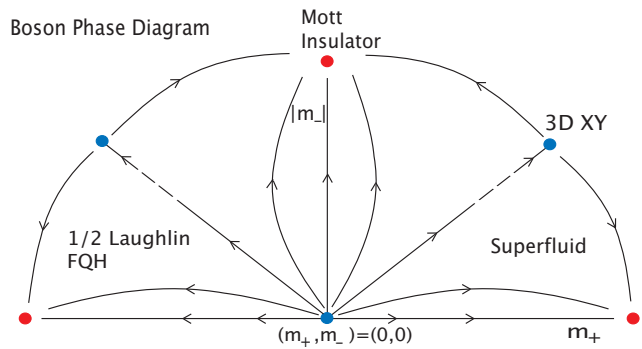}
}
\caption{Proposed phase diagram and renormalization-group flows including the Mott insulator, superfluid, and 
$\nu = 1/2$ Laughlin FQH state, for fixed average particle number.\cite{BM1293} We have defined
$m_\pm \equiv m_1 \pm m_2 $ 
(see eq.~\ref{CSDLag} with $N_f =2$);
$m_-$ can be viewed as a symmetry-breaking field, so the direct transition between the FQH state and the SF can occur if the symmetry is preserved. 
The red points on the horizontal and vertical axes indicate the three stable phases, while the blue points at the origin 
and the diagonals indicate the unstable critical fixed points. 
\label{bosonPhaseDiag}
}
\end{figure}

From this framework, we can now understand transitions out of the CFL state and ultimately into the Fermi liquid by
considering transitions in the boson sector between the $\nu =1/2$ Laughlin FQH state, the Mott insulator, and the superfluid. 
In \Ref{BM1293}, we found that in the absence of any special symmetries and for fixed particle number,
there are continuous transitions between the bosonic Laughlin FQH state and the Mott insulator, and between the
boson Mott insulator and the superfluid. However, certain spatial symmetries, such as inversion, can stabilize a direct continuous transition
between the FQH state and the superfluid. These critical points are described by massless Dirac fermions coupled to 
a U(1) CS gauge field:
\begin{align}
\label{CSDLag}
\mathcal{L}_{N_f, k} = \frac{N_f k}{4\pi} \epsilon^{\mu \nu \lambda} a_\mu \partial_\nu a_\lambda + 
\sum_{i=1}^{N_f} [\bar{\psi}_i \gamma^\mu D_\mu \psi_i + m_i \bar{\psi}_i \psi_i].
\end{align}
It was found that the MI-SF transition is described 
as $m_1 \to 0^+$ in $\mathcal{L}_{1,1/2}$, the FQH-MI transition
is described by $m_1 \to 0^+$ in  $\mathcal{L}_{1,3/2}$, and the FQH-SF transition is described by 
$m_1 = m_2 \equiv m \to 0^+$ in $\mathcal{L}_{2,1/2}$ (see Fig.~\ref{bosonPhaseDiag})\cite{CF9349,BM1293}.
Note that for the above theory to have a well-defined lattice regularization, $N_f k$ must be an integer
when $N_f$ is even, and half-integer when $N_f$ is odd. The critical theory at $m = 0$ may be modified in
the presence of long-range interactions, depending on the value of $k$ \cite{YS9809}. It was found that in 
the large $N_f$ limit, for $k > k_{c1}$, Coulomb interactions are 
marginally irrelevant, while for $k_{c2} < k < k_{c1}$, Coulomb interactions are relevant and cause a flow to a different stable
fixed point with dynamic critical exponent $z =1$ and correlation length exponent $\nu > 1$. 
For $k < k_{c2}$, Coulomb interactions are relevant and flow to strong coupling. 
It was found that $k_{c1} \approx 0.35$ and $k_{c2} \approx 0.28$. Thus for the cases of interest here, it is possible that
Coulomb interactions will either be marginally irrelevant or flow to the controlled fixed point with $z=1$ and $\nu > 1$. 

We note that since time-reversal symmetry is strongly broken in the situation under consideration, a possibility is
that at the FQH - SF critical point, the initial transition out of the FQH state is into a vortex state of the superfluid. 
If the vortices form a vortex lattice, then both translation and $U(1)$ charge symmetry are broken at the 
transitions. We expect that such a scenario would be multicritical and would not be described by (\ref{CSDLag});
any translation symmetry breaking should generically occur away from the FQH to SF critical point. 

Given the above critical points between the different bosonic states, it is possible that the electron
system will also undergo continuous transitions between the CFL, gapless MI, and FL states
as the bosonic sector of the theory undergoes transitions between the $\nu  =1/2$ Laughlin state,
the Mott insulator, and the superfluid. 

In order to analyze whether the resulting transitions of the electron system are continuous, we must
analyze the coupling of the bosons to the fermions and the gauge field. The effective
theory takes the form
\begin{align}
\mathcal{L} = \mathcal{L}_b + \mathcal{L}_f + \mathcal{L}_a + \mathcal{L}_{bf},
\end{align}
where $\mathcal{L}_b$ is the action for the boson sector, $\mathcal{L}_f$ is the action
for the $f$ fermions, which fill a Fermi surface, and 
\begin{align}
\mathcal{L}_a = \frac{1}{g^2} (\nabla \times a)^2,
\end{align}
and $\mathcal{L}_{bf}$ contains direct boson-fermion couplings; $g$ is a phenomenological parameter
in the effective theory. First, we will consider the transitions at the mean-field level, where we ignore fluctuations of the 
emergent $U(1)$ gauge field $a$. In the absence of $\mathcal{L}_{bf}$, then, the boson critical point will be described
by $\mathcal{L}_{N_f, k}$, for a suitable choice of $N_f$ and $k$. Let us consider the possible effects
of $\mathcal{L}_{bf}$. An operator $O$ from the boson sector can couple to the particle-hole continuum of the $f$ Fermi surface
at low momenta. We can use the arguments of \Ref{SM0217} to see whether such a coupling is relevant at the
boson critical point. Integrating out the $f$ fermions gives rise to a perturbation
\begin{align}
v \int \frac{|\omega|}{q} |O(q,\omega)|^2,
\end{align}
for $\omega \ll q$. The most relevant operator is expected to be $O = b^\dagger b$, which 
has scaling dimension $3-1/\nu$ at the boson critical point. 
For this operator, $v$ is therefore irrelevant at the boson critical point if $\nu > 2/3$, and relevant if $\nu < 2/3$. 

From the large $N_f$ expansion of $\mathcal{L}_{N_f,k}$,\cite{CF9349,YS9809}
\begin{align}
\nu^{-1} = 1 + \frac{512 \phi (1 - 2\phi)}{3 \pi^2 (1+\phi)^3} \frac{1}{N_f} + O(1/N_f^2),
\end{align}
where $\phi = (\theta/16)^2$ and $\theta = 2\pi/k$. By comparing the value of $\nu_{1,1/2}$ with the
known 3D XY value, we conclude that the large $N_f$ expansion is unreliable for $N_f = 1$. For $N_f = 2$,
we expect the large $N_f$ expansion to be more accurate, and we find that to $O(1/N_f^2)$, $\nu_{2,1/2} = 1.4182 > 2/3$. 
Thus in the absence of gauge fluctuations and for short-ranged interactions, we see that the direct symmetry-protected CFL to FL transition
is continuous according to the leading order $1/N_f$ approximation. Since we don't have accurate 
estimates of $\nu_{1,3/2}$, we cannot conclude that the CFL to GMI transition is also continuous. 
The GMI to FL transition is continuous: there, the boson transition is in the 3D XY universality class, for which $\nu_{3D XY} > 2/3$.
Since both the GMI-FL and the CFL - FL are continuous, in what follows, we will consider the possibility that the CFL to 
GMI transition is also continuous and study the phenomenology of such a transition, along with that of the CFL - FL transition. 

In the presence of the long-ranged Coulomb interactions, there are several possibilities. If the bosonic sector flows to the
new fixed points found in \Ref{YS9809}, then $\nu > 1$, and so $v$ will be an irrelevant perturbation in all cases. Alternatively, if the long-ranged
Coulomb interactions are marginally irrelevant, then scaling functions will receive logarithmic corrections. 

\begin{figure}
\centerline{
\includegraphics[width=3.3in]{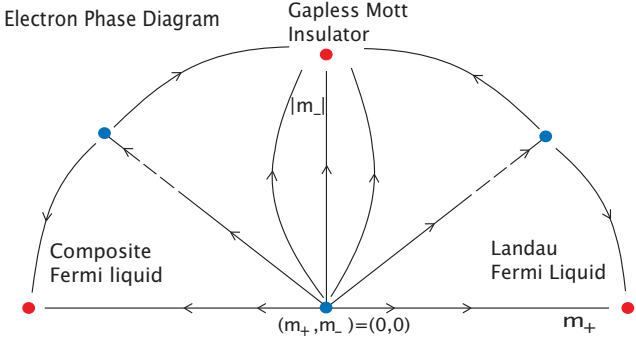}
}
\caption{
The red points on the horizontal and vertical axes indicate the three stable phases, while the blue points at the origin 
and the diagonals indicate the unstable critical fixed points. 
\label{elPhaseDiag}
}
\end{figure}

\subsection{Effect of gauge fluctuations}

In \Ref{S0809}, the effect of gauge fluctuations has been studied in the case of the gapless MI to
FL transition, where the bosons are undergoing a 3D XY transition and time-reversal symmetry is preserved. 
The phenomenology here is similar; the main non-trivial differences arise from the 
non-zero Hall conductivity of the boson sector at the critical points. 
We note that the analysis of the gauge fluctuations relies on results obtained from RPA calculations, whose 
validity in large $N$ expansions is discussed in 
Refs.~\onlinecite{L0902, MS1007, MM1004, NW9459}.


First, consider integrating out all matter fields. The resulting action for the gauge field, to quadratic order in $a$, is
\begin{align}
S_{eff}[a] = \frac{1}{2}\int a_\mu (\Pi_{\mu \nu}^{f} + \Pi_{\mu \nu}^b) a_\nu.
\end{align}
Since the boson critical point has dynamical scaling exponent $z = 1$
and rotation invariance,
the general form of the boson polarization tensor $\Pi_{\mu\nu}^b$ at $T = 0$ is
\begin{align}
\Pi_{\mu \nu}^b = (\delta_{\mu \nu} - \frac{k_\mu k_\nu}{k^2}) \Pi_+^b(k) + \epsilon^{\mu \nu \lambda} k_\lambda \Pi_-^b(k),
\end{align}
where $\Pi_+^b$ and $\Pi_-^b$ are functions of $k \equiv \sqrt{q^2 + \omega^2}$,
and $k_\mu$ is a three-component vector $(\omega, \vec{q})$, where $\vec{q}$ is the wave vector.  
At the boson critical point, $\Pi_-^b(k) \sim \mathcal{O}(k^0)$ and $\Pi_+(k) \sim k$.\cite{CF9349}

In Coulomb gauge, $\partial \cdot a = 0$, the gauge field consists of $(a_0, a_{\perp})$, where
$a_{\perp}$ is the component transverse to the wave vector. The polarization tensors can then be written
as 2$\times$2 matrices for the time and transverse components:
\begin{align}
\label{bPol}
\Pi^b = \left(\begin{matrix} 
\frac{q^2}{k} f_1\(\xi k\) & - q g\(\xi k \) \\
q g\(\xi k\) & k f_2\(\xi k\)
\end{matrix} \right) ~,
\end{align}
where $\xi$ is the correlation length of the boson critical point. On the FQH and Mott
insulating side of the critical points, 
\begin{align}
f_i(x \rightarrow \infty) &\rightarrow f_0,
\nonumber \\
f_i(x \rightarrow 0) &\rightarrow x,
\nonumber \\
g(x \rightarrow \infty) &\rightarrow \sigma_{xy}^c,
\end{align}
where $f_0$ is a constant, and $\sigma_{xy}^c$ is a constant setting the Hall conductivity at the critical point. 
In the FQH state, $g(x \rightarrow 0) \rightarrow \frac{1}{2}$; in the Mott insulating state,
$g(x \rightarrow 0) \rightarrow 0$. On the superfluid side of the critical points, 
\begin{align}
f_i(x \rightarrow \infty) &\rightarrow f_0,
\nonumber \\
f_i(x \rightarrow 0) &\rightarrow 1/x,
\nonumber \\
g(x \rightarrow \infty) &\rightarrow \sigma_{xy}^c,
\nonumber \\
g(x \rightarrow 0) &\rightarrow \sigma_{xy}^s,
\end{align}
where we allow for the possibility of a non-zero Hall conductivity $\sigma_{xy}^s$ in the superfluid phase. 
The fermion polarization $\Pi_f$ is, for $q \ll k_F$ and $\omega \ll v_F q$,
\begin{align}
\label{eq:RPAf}
\Pi_f = \left( \begin{matrix}
\kappa_f & \Pi_{f;xy}(q,\omega) \\
\Pi_{f;yx}(q,\omega) & \frac{k_0|\omega|}{q} + \chi_d q^2 \\
\end{matrix} \right),
\end{align}
where $\kappa_f$, $\chi_d$, and $k_0$ are constants. We note that in general, since time-reversal symmetry is broken,
the $f$ fermions may have a Hall conductance $\sigma_{xy}^f$, so $\Pi_f$ can also have off-diagonal components: 
$\Pi_{f;xy}(q,\omega = 0) =\Pi_{f;yx}^*(-q,\omega=0) = -q \sigma_{xy}^f$. 

The arguments of \Ref{S0809} imply that the gauge fluctuations also do not modify the boson critical point. 
This can be understood by observing that the gauge fluctuations, using the RPA propagator from (\ref{eq:RPAf}), 
only lead to analytic corrections to the boson self-energy at low $\omega$, $q$, and therefore do not modify
the critical singularities coming from the boson sector. Alternatively, the $\omega/q$ term acts like a Higgs 
mass for the transverse gauge fluctuations in the boson sector, since $\omega$ and $q$ scale the same way 
at the boson critical point.

Using (\ref{ioffeLarkin}), (\ref{bPol}), and (\ref{eq:RPAf}), we can obtain the compressibility $\kappa_e$
at zero temperature close to the transition. The inverse electron polarizability satisfies
\begin{align}
\Pi_e^{-1}(q, \omega = 0, T = 0) = \left(\begin{matrix}
\frac{\chi_d q^2}{|\Pi_f|} + \frac{q f_2}{|\Pi_b|} & \frac{q \sigma_{xy}^f}{|\Pi_f|} + \frac{qg}{|\Pi_b|}\\
-\frac{q \sigma_{xy}^f}{|\Pi_f|} - \frac{qg}{|\Pi_b|} & \frac{\kappa_f}{|\Pi_f|} + \frac{q f_1}{|\Pi_b|}
\end{matrix} \right),
\end{align}
where $|\Pi_f| \propto q^2$ and $|\Pi_b| = q^2(f_1 f_2 + g^2)$ are the determinants of $\Pi_f$ and $\Pi_b$.
We are interested in the case where $q$ and $q\xi$ are small, while $\xi$ diverges. 
On the CFL side, in this limit $|\Pi_b| \propto q^2$, $f_i \sim q \xi$, and $g \sim const.$.
On the LFL side, $f_i\sim 1/q\xi$, $g \sim const.$, and $|\Pi_b| \sim 1/\xi^2$.
Therefore for small $q$ and $q \xi$, on the CFL and LFL side, the dominant terms are
\begin{align}
\Pi_e^{-1}(q, \omega = 0, T = 0) \approx \left(\begin{matrix}
\frac{q f_2}{|\Pi_b|} & \frac{\alpha}{q}\\
-\frac{\alpha}{q} & \frac{\kappa_f}{|\Pi_f|} 
\end{matrix} \right),
\end{align}
for some constant $\alpha$, and so
\begin{align}
\Pi_{e;00}(q,\omega =0, T=0) \sim \left(\frac{q f_2 }{|\Pi_b|} + \frac{\alpha^2 |\Pi_f|}{q^2 \kappa_f} \right)^{-1}.
\end{align}
As $\xi \rightarrow \infty$, the first term dominates and we get 
$\kappa_e = \lim_{q \rightarrow 0} |\Pi_b/qf_2| \sim 1/\xi$.

Therefore, as the critical point is approached from either the CFL side or the LFL side, 
we find that the compressibility $\kappa_e$ 
at zero temperature close to the transition is dominated by the correlation length of the boson sector:
\begin{align}
\kappa_e \sim 1/\xi.
\end{align}
On the LFL side, $1/\xi \sim \rho_s$, while on the CFL side, $1/\xi \sim \Delta_b$, where $\Delta_b$ is the gap
of the boson sector and $\rho_s$ is the superfluid density.

\subsection{Transport and thermodynamics at the critical points}

From the Ioffe-Larkin composition rule (\ref{ioffeLarkin}), the inverse polarizability of the electrons is given
by the sum of the inverse polarizabilities of $f$ and $b$.
The bosonic sector has a Hall conductance, $\Pi_{b, 0i} \neq 0$  and therefore the inverse compressibilities
and resistivities of $b$ and $f$ do not separately add to yield the electron inverse compressibility and resistivity.
This implies that while the boson resistivity exhibits a universal jump,\cite{FG9087} there is also a jump in the resistivity of the electron system;
in contrast to the time-reversal invariant case, the resistivity jump appears to be non-universal because of the off-diagonal terms in the polarization tensors.

At finite temperature and zero frequency, by scaling we have
\begin{align}
\label{polfiniteT}
\Pi_{b; \mu \nu}(q,\omega = 0, T) = T f_{\mu \nu}(q/T),
\end{align}
where $f_{\mu\nu}$ is a scaling function satisfying
\begin{align}
\label{polineq}
0 <\lim_{q \rightarrow 0} |f_{\mu \nu}(q/T)| < \infty.
\end{align}
(\ref{polineq}) can be understood by observing that the polarization function $\Pi_{\mu \nu}(q, \omega, T=0)$
vanishes as $q,\omega \rightarrow 0$; at finite temperature, the current-current correlation functions should be
more short-ranged in real space, and therefore should continue to vanish as $q, \omega \rightarrow 0$. 
Furthermore, generically $|f_{\mu \nu}(0)| > 0$ as there is no symmetry forcing it to be zero. 

The fermion polarization at $\omega = 0$ and small $q$ is temperature independent
and given by (\ref{eq:RPAf}).

Using (\ref{ioffeLarkin}), (\ref{polineq}), (\ref{polfiniteT}) (\ref{eq:RPAf}), the inverse electron polarizability is:
\begin{align}
\Pi_e^{-1}(q,\omega = 0, T) = \frac{1}{T} f^{-1} + \frac{1}{|\Pi_f|} 
\left( \begin{matrix}
\chi_d q^2 & q \sigma_{xy}^f\\
-q \sigma_{xy}^f & \kappa_f
\end{matrix} \right),
\end{align}
where the fermion polarization determinant $|\Pi_f| \propto q^2$.
In the limit $q/T\rightarrow 0$ and $T \rightarrow 0$, the dominant terms are:
\begin{align}
\Pi_e^{-1} \sim \left(\begin{matrix}
\frac{1}{T} (f^{-1})_{00} & \frac{q \sigma_{xy}^f}{|\Pi_f|} \\
\frac{-q \sigma_{xy}^f}{|\Pi_f|} & \frac{\kappa_f}{|\Pi_f|}
\end{matrix} \right).
\end{align}
Therefore:
\begin{align}
\Pi_{e;00} \sim \left( \frac{1}{T} (f^{-1})_{00}  + \frac{q^2 (\sigma_{xy}^f)^2}{\kappa_f |\Pi_f|} \right)^{-1}.
\end{align}
The first term dominates at low temperatures, from which we find that the compressibility is
$\kappa_e(T) \equiv \lim_{q \rightarrow 0} \lim_{\omega \rightarrow 0}  \Pi_{e;00}(q, \omega, T)$,
\begin{align}
\kappa_e \sim T + \mathcal{O}(T^2).
\end{align}
A remarkable prediction of the above scaling of the compressibility is that
the critical point between the composite Fermi liquid and Landau Fermi liquid is incompressible
at zero temperature, even though both the CFL and FL are compressible states. 

At the critical point, the gauge propagator in RPA is equivalent to that considered in the Halperin-Lee-Read
theory of CFL with long-range interactions. Since the specific heat of the boson sector is $C_v \sim T^2$,
we therefore expect the specific heat to be dominated by the contribution of the fermion-gauge system:
\begin{align}
C_v \sim T \ln (\frac{1}{T}).
\end{align}

The expectations for transport are very similar to those of \Ref{S0809,WG1209}.

\subsection{Crossover out of criticality}

An important signature of this class of transitions is that while the boson 
critical point has a quantum critical region associated with some crossover temperature 
scale $T^* \sim 1/\xi$, the gauge propagator is not significantly modified until a lower temperature scale, 
$T^{**} \sim \frac{1}{\xi^2 ck_0}$, where $c$ is the characteristic velocity of excitations at the boson critical
point (previously we had set $c = 1$), and $k_0$ is defined in (\ref{eq:RPAf}).\cite{S0809} 
Recall $\xi \sim 1/\Delta_b$, where $\Delta_b$ is the boson gap when the bosons are in the gapped
FQH or MI states, and $\xi \sim 1/\rho_s$ when the bosons are in the superfluid state.
This implies the existence of two finite temperature crossover regimes, each with distinctive properties. 


In the crossover regime $T^{**} < T < T^*$, the gauge propagator still takes the form that it did
in the quantum critical regime of the boson critical point. 
On the composite Fermi liquid side, this means that the theory formally is
similar to the composite Fermi liquid theory with long-range Coulomb interactions and we expect
the physics of this theory to dominate this crossover regime. In such a situation, the composite fermions
form a ``marginal'' Fermi liquid. If the problem does have long-range Coulomb interactions to begin 
with, there will be no significant modification of the physics on the CFL side as one crosses $T^{**}$. 
On the other hand, if the Coulomb interaction is screened by a gate, then $T^{**}$ 
will mark a crossover between shorter-ranged and effectively long-ranged interactions. 
The fact that this crossover scale $T^{**}$ can effectively be made to disappear and reappear with a
screening gate is a non-trivial prediction of this theory. 

The compressibility below $T^*$ is dominated by the bosons and will be a temperature-independent constant
$\kappa_e \sim 1/\xi$.

\begin{figure}[t]
\centerline{
\includegraphics[width=3.3in]{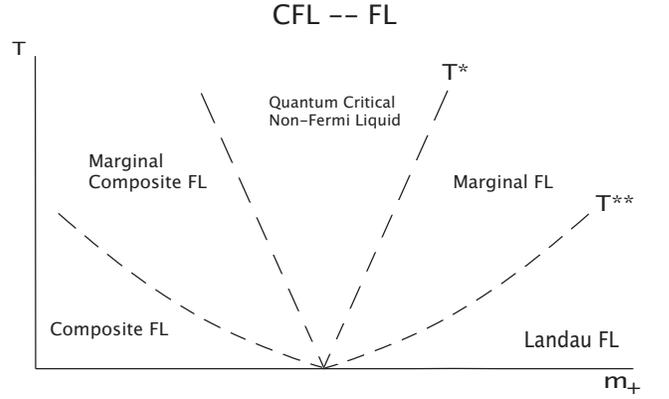}
}
\caption{
\label{CFLFL}
Schematic phase diagram showing finite temperature crossover regimes between the
CFL and LFL. For $T > T^*$, the bosons are in their quantum critical regime. For $T^{**} < T < T^*$,
on the CFL side the system is a marginal Fermi liquid of composite fermions; on the
LFL side, the system is a marginal Fermi liquid of electrons.
}
\end{figure}
\begin{figure}[t]
\centerline{
\includegraphics[width=3.3in]{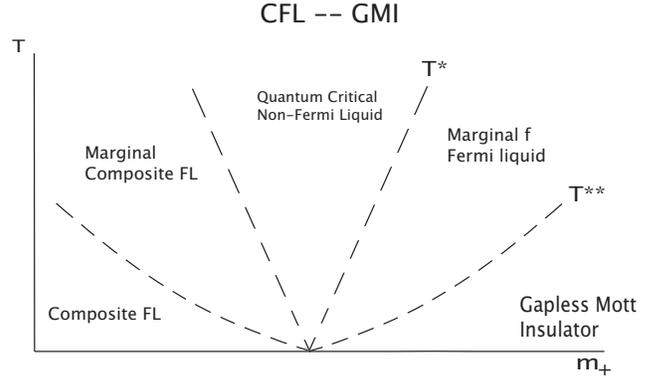}
}
\caption{
\label{CFLGMI}
Schematic phase diagram showing finite temperature crossover regimes between the
CFL and GMI. For $T > T^*$, the bosons are in their quantum critical regime. For $T^{**} < T < T^*$,
on the CFL side the system is a marginal Fermi liquid of composite fermions; on the
GMI side, the system is a marginal Fermi liquid of the emergent $f$-fermions.
}
\end{figure}

The crossover behaviors on the LFL and GMI sides of 
these transitions are spinless analogs of the cases studied in \Ref{S0809}; we will
briefly review a few of the key features here;
the results are summarized in Figs.~\ref{CFLFL} and \ref{CFLGMI} respectively.


On the GMI side, in the second quantum critical regime, $T^{**} < T < T^*$, the specific heat is still behaving as
\begin{align}
C_v \sim T \ln T.
\end{align}
The system can be viewed as a marginal Fermi liquid of the neutral $f$ excitations,
although the single-particle properties of $f$ are not gauge invariant and cannot be directly measured.
The thermal conductivity in this regime behaves as $K/T \sim 1/T$.

At the lowest temperatures away from the critical point, $T < T^{**}$, for short-range interactions the
system has a specific heat
\begin{align}
C_v \sim T^{2/3},
\end{align}
and the electron single-particle Green's function decays exponentially in frequency, indicating an
exponential suppression of the electronic density of states. In this regime, the thermal 
conductivity behaves as $K/T \sim 1/T^{2/3}$.


On the Landau FL side, at $T= 0$ and $\frac{1}{\xi^2 ck_0} < \omega < 1/\xi$ the electron Green's function 
has the marginal Fermi liquid form,\cite{VL8996} with a self-energy
\begin{align}
\Sigma(\omega) \sim \omega \ln i\omega. 
\end{align}
In the crossover temperature regime $T^{**} < T < T^*$, the specific heat behaves as
\begin{align}
C_v \sim T \ln T,
\end{align}
while the thermal conductivity is
\begin{align}
K/T \sim T.
\end{align}
The compressibility below $T^*$ is dominated by the bosons and is a temperature-independent 
constant, $\kappa_e \sim 1/\xi$. As the transition is approached from the Landau FL side, the 
quasiparticle effective masses and Landau parameters are expected to diverge in the same manner
as described in \Ref{S0809}.

\section{Onset of fermion pairing}
\label{pairedSec}

Another possible neighbor of
the composite Fermi liquid state of the half-filled Landau level
is a paired superconducting state of the composite fermions. For example, when the composite fermions are paired into
a topologically non-trivial $p_x + ip_y$ superconducting state, the result is a description of the non-Abelian Moore-Read Pfaffian state, which
is a candidate state for the plateau at $\nu  = 5/2$ in GaAs quantum wells\cite{MR9162, RG0067}. The properties of the transition 
from CFL to the Moore-Read state are the subject of ongoing investigations by others. 

In the context of this paper, it is possible in principle that the pairing transition of the composite fermions occurs
before the transition out of the composite Fermi liquid state and into the gapless Mott insulator or the Landau Fermi liquid, depending on the interactions between the composite fermions. 
In this case, the Landau Fermi liquid is replaced by a paired electronic superconductor, while the gapless Mott insulator is replaced
by a different topologically ordered state, consisting of the fermion condensate coupled to an emergent $Z_2$ gauge field.
For example, 
the Moore-Read Pfaffian
has a topological ground state degeneracy of six on the torus, and the gapless chiral edge theory
consists of the Ising $\times$ U(1) chiral conformal field theory (CFT) with central charge $c = 3/2$. 
When the fermions of the gapless Mott insulator are paired into a topologically non-trivial (weak-paired) $p_x + i p_y$ state, we instead obtain
a topological phase with a torus ground state degeneracy of three, and a $c = 1/2$ Majorana fermion edge state; such 
a state has the ``Ising'' topological order,\cite{NS0883} and is well-known as the non-Abelian state in the ``B phase''
of Kitaev's honeycomb model.\cite{K0602}

While the superconducting state cannot exist in the presence of a strong magnetic field, these transitions may be induced
by starting with the half-filled Landau level and turning on a periodic potential with $2\pi$ flux per plaquette. In the
limit where the superconducting state is obtained, the effect of the magnetic field disappears because
the electrons are confined to paths along the periodic potential and therefore do not feel a net magnetic field, as the phase 
through all closed paths will be a multiple of $2\pi$. 
\begin{figure}[t]
\centerline{
\includegraphics[width=3.3in]{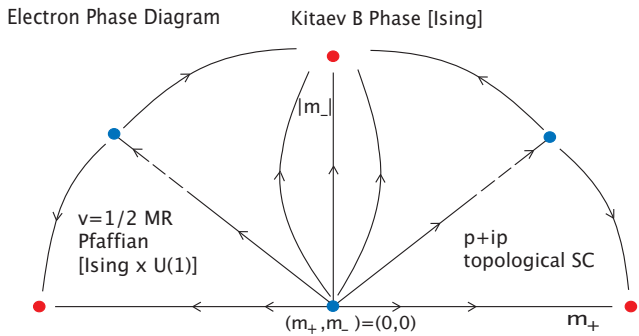}
}
\caption{
\label{naPhaseDiagram}
Schematic phase diagram of the electron system when the fermions are paired into a topologicaly non-trival $p_x+ip_y$
state. The transitions are driven by the bosonic sector undergoing $\nu=1/2$ Laughlin FQH - MI -- SF transitions.
Since the fermions are gapped, the critical theories are the same as for the boson system, given by eq. (\ref{CSDLag}).
}
\end{figure}
In these paired states, the $U(1)$ emergent gauge symmetry is broken to $Z_2$. Since the gauge fluctuations are gapped, at the critical point
they can be integrated out to obtain short-range interactions among the bosons. At the boson critical points, these short-ranged interactions are
irrelevant. Therefore, the critical theories separating these gapped states are the same critical theories (\ref{CSDLag}) that separate
the $\nu  =1/2$ Laughlin state, the Mott insulator, and the superfluid of bosons. Furthermore, since
the gauge fluctuations are gapped, one obtains a single crossover temperature scale associated with the quantum critical points. 
Interestingly, since the fermions remain in a gapped state throughout, the electron operator is not a scaling operator 
at these quantum phase transitions and has exponentially decaying correlations at the quantum critical points.

The existence of a continuous bandwidth-tuned transition between the $\nu = 1/2$ Moore-Read Pfaffian state 
and the Kitaev B (Ising) phase, and the symmetry-protected quantum critical point between the $\nu = 1/2$ Moore-Read Pfaffian and the
$p_x + ip_y$ electronic superconductor are highly non-trivial predictions of the theories presented here. 


\section{Discussion}
\label{summarySec}

In this paper, we developed a theory of transitions out of the composite Fermi liquid and ultimately
into a Landau Fermi liquid. These transitions can be induced by tuning the bandwidth of the partially
filled band. In the Landau level problem, this can be done by using an external periodic potential; if the composite
Fermi liquid is instead realized in a partially filled Chern band, the bandwidth can be tuned with pressure.  

We found that generically, the transition to the Landau Fermi liquid occurs through an intermediate 
gapless Mott insulating phase, with a Fermi surface of neutral fermions. In the presence of certain 
spatial symmetries, such as inversion, we found a direct continuous transition between the composite 
Fermi liquid and the Landau Fermi liquid, providing a highly non-trivial example 
of a quantum critical point between a fractionalized non-Fermi liquid and a conventional Fermi liquid. 

In order to establish the above results, we had to depart from the conventional understanding of
the composite Fermi liquid, where the electron is viewed as a fermion attached to flux quanta. Instead, 
the electron fractionalizes into a boson $b$ and a fermion $f$. When $b$ forms a $\nu = 1/2$ incompressible Laughlin
state, at low energies $b$ is effectively created by inserting 2 flux quanta, so the electron can be viewed
as 2 flux quanta attached to the fermions, which leads to the composite Fermi liquid description. 
Another natural possibility is that depending on the nature of the interactions and the bandwidth, 
$b$ can form a Mott insulator to minimize the interaction energy. In this case, the state of the electrons is a Mott insulator,
which is gapless when $f$ has a Fermi surface. Finally, when the bandwidth is large enough,
the bosons will condense into a superfluid, leading to a description of the Landau Fermi liquid.

The direct CFL -- FL transition found here is not stable to disorder, as it relies on the presence of spatial symmetries
that can protect a pair of Dirac cones. Nevertheless, in the presence of weak disorder and small but finite $T \neq 0$, the properties of the direct CFL -- FL
quantum critical point can determine the physics. Despite the fact that weak disorder will ultimately render the direct transition
unstable at the lowest temperatures, the relevance of this depends on the relative strength of disorder, $W$, and temperature $T$. For 
$W \ll T$, this ultimate instability is of no measurable consequence. 

The situation is somewhat less clear for the CFL - GMI transition, because there is currently no reliable
estimate for its correlation length exponent $\nu$. While the boson FQH - MI transition does not require any spatial
symmetries, it is expected to be unstable to disorder if $\nu < 1$; in this case, the theory will flow to a different
critical point, with $\nu \geq 1$, that is stable to disorder. Whether the putative clean critical point or the
disordered one is the relevant one for describing the physics depends again on $W/T$. 

Interestingly, the currently understood direct transition\cite{BM1293} between the Laughlin state and the superfluid exists only
for the $\nu  =1/2$ Laughlin state.\footnote{For other boson FQH states, the direct transition to the superfluid appears to
be highly multi-critical.} This implies that the direct continuous transition between the CFL
and the Fermi liquid can currently only be understood at half-filling. For other more general CFL states at other filling fractions,
the route to the Landau Fermi liquid appears to always be separated by an intervening gapless Mott insulator,
although there can be a multi-critical point directly separating the CFL and LFL. This difference in the properties
between $\nu=1/2$ and $\nu=1/2m$ for $m > 1$ appears to form a counterexample to the law of corresponding states\cite{KL9223}
that was suggested for previously understood FQH transitions.

The experimentally observable phenomenology of the transitions includes the existence of two crossover 
temperature scales and resistivity jumps at the transition, and a vanishing compressibility at the critical points,
similar to the $U(1)$ spin liquid Mott transition \cite{S0809}. We find that the composite Fermi liquid 
provides another experimentally promising venue where the physics of 
such slave-particle gauge theory transitions in the presence of a Fermi surface, and novel exotic fractionalized phases, can be studied. 

The theory developed here assumes that the system does not hit a first-order transition out of the composite 
Fermi liquid. The validity of this assumption depends on microscopic details of the interactions.
If there is a first-order transition, then both disorder and/or long-range interactions can render the first-order transition to
be continuous.\cite{SK0671} Such continuous transitions would exhibit completely different physics from that studied in this paper
and would require a completely different theory. 

\vskip.2in
\it Acknowledgements \rm -- We thank N.~Bonesteel, S. Kivelson, S. Raghu, S. Parameswaran, 
B. Swingle, Cenke Xu, M.P.A. Fisher, and especially T. Senthil for helpful discussions. We also acknowledge the KITP programs 
``Topological Insulators and Superconductors,'' and ``Holographic Duality and Condensed Matter Physics''
for hospitality while part of this work was done. This work was supported by a Simons Fellowship (MB)
and by the U.S. Department of Energy
(D.O.E.) under cooperative research agreement DE-FG0205ER41360,
and by the Alfred P. Sloan Foundation (JM).

\appendix

\bibliography{biblio.bib}

\end{document}